\documentclass[
  aps,
  prx,
  preprint,
  superscriptaddress,
  longbibliography,
  nofootinbib
]{revtex4-2}

\usepackage{amsmath,amssymb,amsfonts,mathtools}
\usepackage{bm}
\usepackage{mathrsfs}
\usepackage{upgreek}
\usepackage{graphicx}
\usepackage{array}[=2016-10-06]
\usepackage[dvipsnames,table]{xcolor}
\usepackage{ytableau}
\usepackage{booktabs,tabularx,longtable}
\usepackage{enumitem}
\usepackage{microtype}
\usepackage[english]{babel}
\usepackage{hyperref}
\hypersetup{
  hidelinks,
  pdftitle={The Young-E(3) Tensor Product Decomposition for Rotation and Permutation Equivariant Cluster Expansions},
}
\ytableausetup{smalltableaux}

\definecolor{ACEBlue}{RGB}{57,105,170}
\definecolor{RoleGreen}{RGB}{52,132,91}
\definecolor{WarmOrange}{RGB}{196,112,45}
\definecolor{DeepRed}{RGB}{145,55,55}
\definecolor{SoftGray}{RGB}{244,245,247}
\definecolor{SkeletonOrange}{RGB}{244,126,32}
\definecolor{SkeletonCyan}{RGB}{0,164,214}
\definecolor{YEBlue}{RGB}{0,114,178}
\definecolor{YEBurntOrange}{RGB}{199,99,32}
\definecolor{YEOrange}{RGB}{230,159,0}
\definecolor{YEPurple}{RGB}{168,105,220}
\definecolor{InactiveGray}{RGB}{158,164,173}

\setlist[itemize]{leftmargin=1.35em,itemsep=2pt,topsep=3pt}
\allowdisplaybreaks
\begin{document}

\title{The Young-E(3) Tensor Product Decomposition for Rotation and Permutation Equivariant Cluster Expansions}
\author{James M. Goff}
\affiliation{Center for Computing Research, Sandia National Laboratories, Albuquerque, New Mexico 87185, USA}
\author{Aidan P. Thompson}
\affiliation{Center for Computing Research, Sandia National Laboratories, Albuquerque, New Mexico 87185, USA}
\date{September 25, 2026}

\begin{abstract}
A generalization of fixed-lattice cluster expansions (CE) and atomic cluster expansions (ACE) is presented that expresses both rotation and permutation symmetries.
By performing Schur--Weyl decomposition in Young subgroup-stabilized carriers, followed by joint coupling of rotation and permutation representations, arbitrary $S_N\times SO(3)$ tensor product carriers are generated.
This allows us to resolve complete, orthonormal joint rotation and permutation-adapted bases for arbitrary tensor product ranks and arbitrary angular and radial tensor product content.
We show that this very general Young--E(3) (YE3T) tensor product basis contains the ACE basis as a subset, corresponding to the special case where permutation symmetry character is restricted to the fully symmetric carrier.
Rather than constructing an overcomplete rotation and permutation-invariant basis and reducing it \textit{a posteriori},
Barthelemy \textit{et al.} recently demonstrated scaling benefits by not constructing an overcomplete basis \cite{barthelemy_efficient_2026}.
YE3T directly produces a complete orthonormal basis without an overcomplete step and rigorously extends to permutation characters beyond permutation-symmetric features.
The YE3T decomposition yields new joint irreducible rotation- and permutation-equivariant basis sets that surpass existing rotation-adapted expansions in both speed and accuracy,
defining a new Pareto front for machine-learned interatomic potentials.
We show that the tunability of the permutation symmetry character makes the basis useful for both atomistic and electronic-structure simulations.
\end{abstract}

\maketitle

\section{Introduction}

The use of rotational equivariance, particularly E(3)-equivariant tensor product representations and basis functions, has resulted in compelling advances in the accuracy of machine-learned atomistic and particle-interaction models\cite{drautz_atomic_2019,batzner_e3-equivariant_2022,batatia_mace_2022,bochkarev_graph_2024}.
Machine-learned interatomic potentials (MLIPs) and electronic-structure models have benefited from forcing models to obey rotational symmetries by construction rather than merely augmenting the training data with rotated and permuted data points.
However, methods that use E(3)-equivariance in conjunction with permutation symmetrization, such as the atomic cluster expansion (ACE\cite{drautz_atomic_2019}) and the equivariant message-passing network potentials (NequIP\cite{batzner_e3-equivariant_2022}, MACE\cite{batatia_mace_2022}, GRACE\cite{bochkarev_graph_2024}), usually only resolve rotational representation content and typically lack formal treatment of permutation character beyond enforcing permutation invariance at some stage.
The resulting tensor product space of permutation-symmetrized rotation products underpinning these methods contains redundant representations and coupling paths when there is repeated content in the tensor product factors \cite{dusson_atomic_2022,goff_permutation-adapted_2024}.
These redundant paths correspond to linearly dependent representations in the hidden feature space of E(3)-equivariant network models and result in overcomplete ACE feature sets.
It is typically the case in ACE-like methods (linear atomic cluster expansions and equivariant neural network potentials) that the rotation and permutation symmetries are treated on a separate footing.
In this work, we show that a richer, more general family of physical modeling methods may be defined if one treats both permutation and rotation symmetries simultaneously.

\begin{figure}[htbp]
\color{black}
\centering
\includegraphics[width=0.76\linewidth]{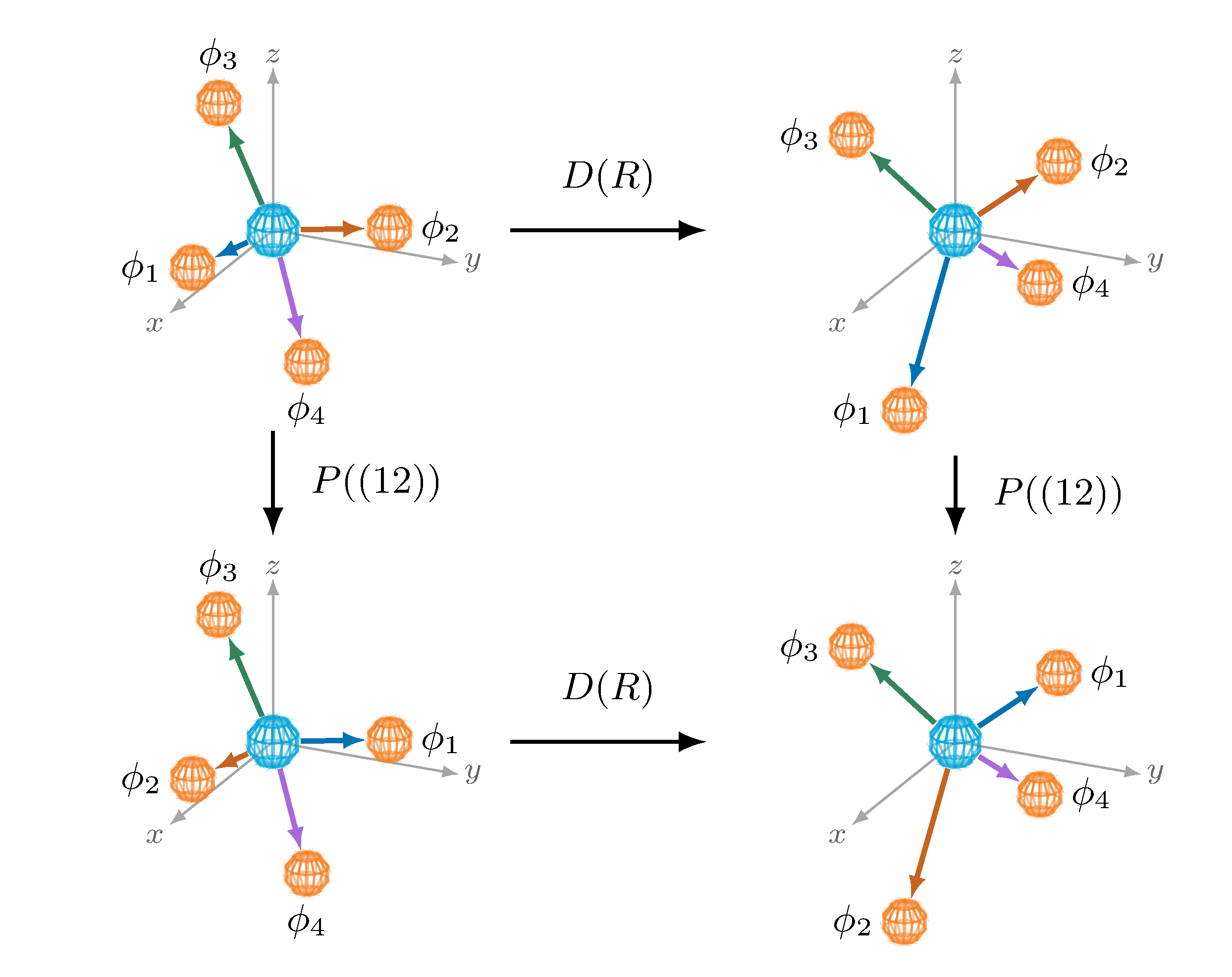}
\caption{
When acting on a tensor product (cluster) interaction, rotations, denoted by $D(R)$ from left to right, and permutations, denoted by $P(\sigma)$ from top to bottom, commute.
In rotation/permutation-equivariant modeling, rotations and permutations of a physical particle system can be viewed in terms of rotations and permutations of interactions.}
\label{fig:intro_commuting_square}
\end{figure}

The choice of applying E(3) equivariance in models, particularly for MLIPs, is often justified because the outputs need to be invariant with respect to rotations and translations (e.g., a rotation of a system should not change the energy) \cite{drautz_atomic_2019,geiger_e3nn_2022}.
Permutation invariance is enforced at early stages for similar reasons \cite{drautz_atomic_2019,zaheer_deep_2017}.
Doing this early throws out permutation information and produces the overcomplete feature sets mentioned above.
To avoid this, we begin by clearly defining rotations, permutations, and their joint actions on general (atomic) cluster expansion bases. We also draw connections to symmetry actions of fixed-lattice cluster expansion models.
These general cluster expansion basis functions are tensor product functions that are used to approximate local physical properties. In this work, basis functions are only forced to respect physical symmetries when necessary.
We show that the permutation action may be viewed as permuting tensor factors or, in other words, permuting the functions of $r_{ab}$ rather than the $r_{ab}$ themselves. Joint rotation/permutation operations on a star cluster interaction are illustrated in Fig.~\ref{fig:intro_commuting_square}.
We show that rotations and permutations can be treated simultaneously using Schur--Weyl decomposition, ultimately leading to irreducible joint rotation- and permutation-equivariant basis sets. We also show that the ACE basis corresponds to the special case of fully symmetric permutation character, i.e., permutation invariance.

Multiple applications of permutation symmetries in machine learning for atomistic and electronic systems would benefit from a formal treatment of both rotations and permutations.
Examples include equivariant machine-learned interatomic potentials and equivariant Hamiltonian learning for quantum systems \cite{batzner_e3-equivariant_2022,gong_general_2023}.
For fermionic systems, antisymmetric permutation character must also be enforced; ACE wave-function representations provide one example \cite{drautz_atomic_2022}.
These problems motivate treating rotation and permutation representation content together rather than resolving the two symmetries independently.

\begin{figure}[htbp]
\color{black}
\centering
\includegraphics[width=0.92\linewidth]{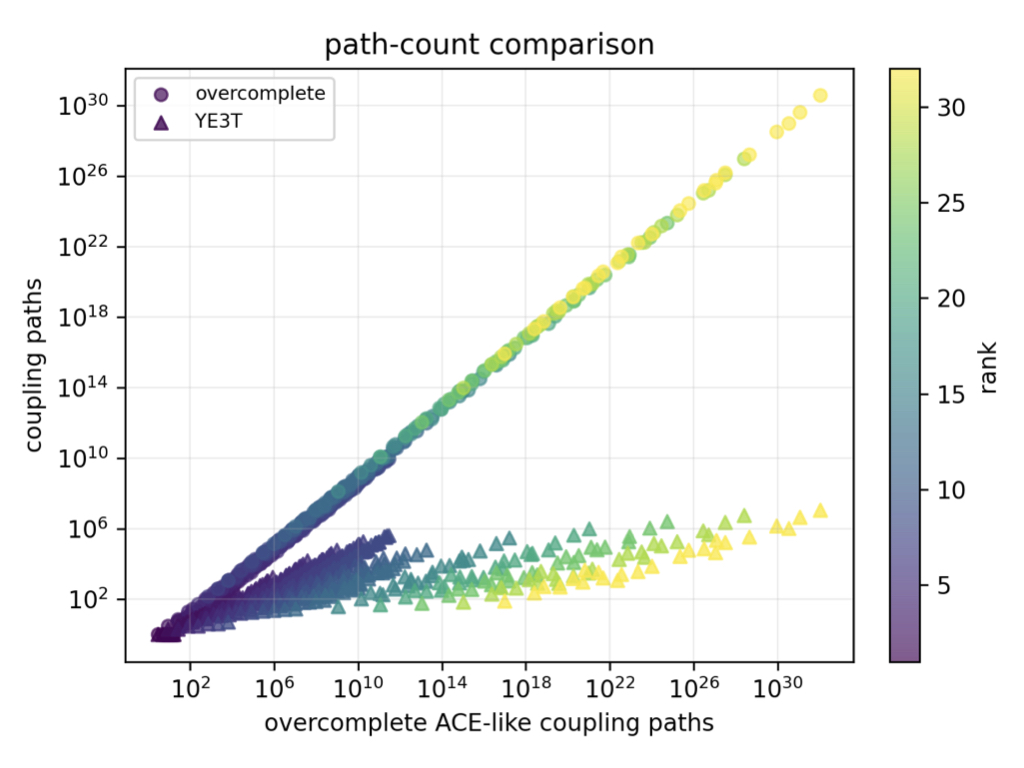}
\caption{Overcomplete tensor products from old constructions of ACE versus irreducible YE3T tensor products, colored by tensor product rank. Previous methods required starting from overcomplete tensor product bases, followed by reduction to the independent ones produced directly by YE3T \cite{dusson_atomic_2022,goff_permutation-adapted_2024}.}
\label{fig:path_count}
\end{figure}

Prior treatments of joint permutation representations in E(3)-equivariant problems, including early implementations underlying ACE and nearly every other E(3)-equivariant MLIP, often treated the overcomplete basis associated with permutation-invariant E(3) tensor products using singular value decomposition (SVD) of descriptor or Gram matrices or generalized angular-momentum recurrence relationships \cite{dusson_atomic_2022,goff_permutation-adapted_2024}.
These require construction of an overcomplete tensor product space followed by numerical reduction.
This overcompleteness makes tensor product expansion models of this type expensive for modest tensor-product ranks between 2 and 9 and renders larger-rank feature construction and modeling completely intractable. This is demonstrated in Fig.~\ref{fig:path_count}.
The YE3T construction in Fig.~\ref{fig:path_count} will be derived later in this paper, but the figure helps illustrate how previous tensor product expansion formalisms for ACE become intractable even at modest rank. This figure was generated for an exhaustive set of radial/angular tensor product channel combinations.

In contrast to earlier methods, YE3T decomposes the tensor product space under the commuting actions of rotations of physical interactions around the center atom and permutations of tensor product factors, which can often be viewed as actions induced by physical-neighbor permutations.
These symmetry operations may be viewed as acting on a joint rotation/permutation carrier space for the tensor product.
In this carrier space, Young subgroup stabilizers and canonical ordering schemes organize repeated channel content blockwise into subsets of repeated channels in which Schur--Weyl decomposition can be used.
Thus, the method is called Young--E(3) tensor (YE3T) decomposition.
Once the tensor-product rank and channel content are specified, YE3T resolves $S_N\times SO(3)$ irreducible representation content and multiplicities.
This is done in a controlled manner, introducing nontrivial permutation character systematically.
Moreover, this produces a complete orthogonal basis corresponding to an irreducible decomposition of $S_N\times SO(3)$ and does so without overcomplete constructions or the need for subsequent numerical reduction.
It is also a more natural and complete description of the physical symmetries relevant for particle interactions compared to methods that only treat the permutation invariant sector.

\begin{figure*}[htbp]
\color{black}
\centering
\includegraphics[width=0.99\textwidth]{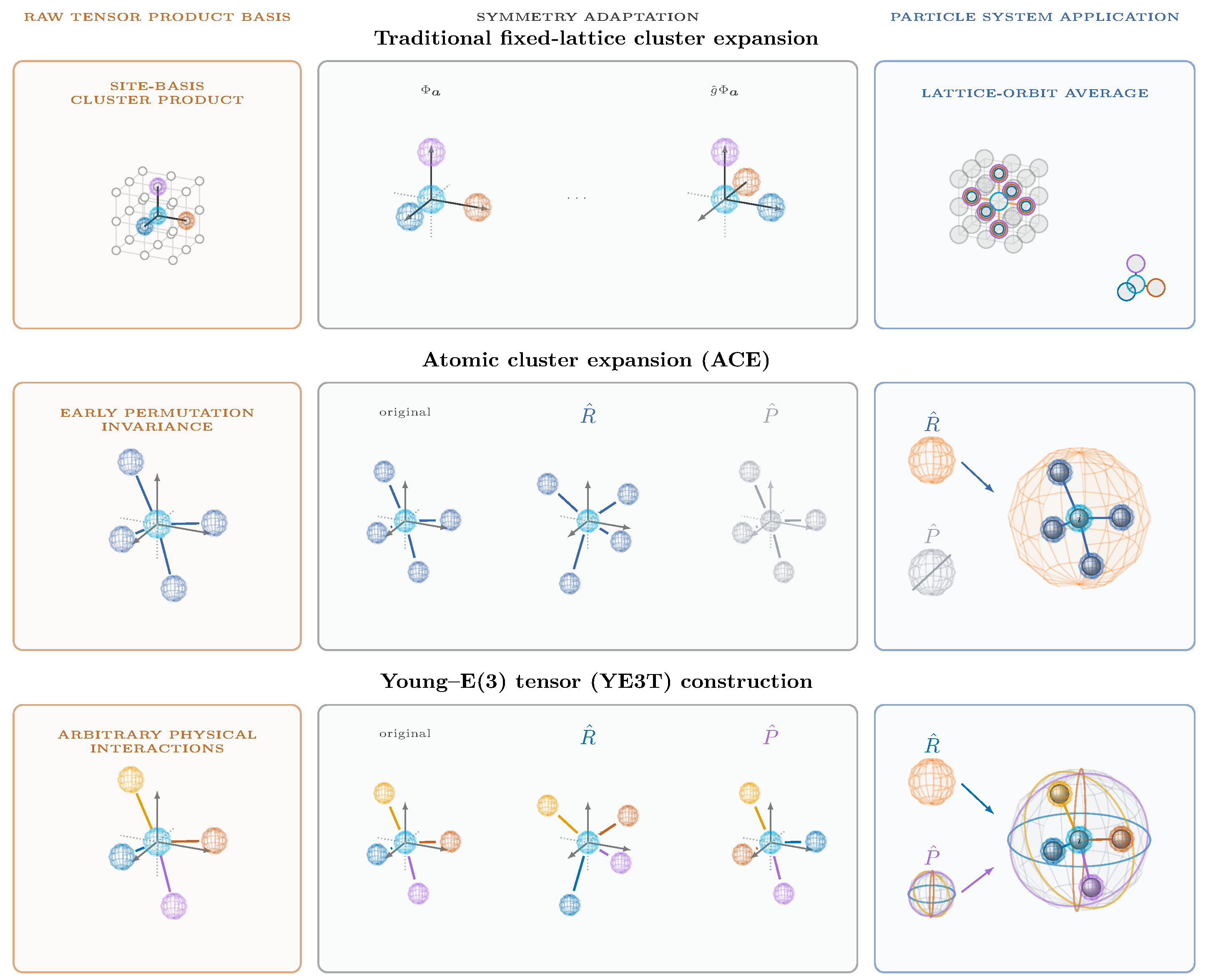}
\caption{
Illustrative evolution of symmetry-adapted tensor product expansion models. The original lattice CE (top) is adapted to the lattice point-group cluster orbit (operations $\hat{g} \in \mathcal{O}$) \cite{sanchez_generalized_1984}.
ACE (middle) is adapted to rotations, but permutation information is averaged out \cite{drautz_atomic_2019}. Gray permutation operations illustrate that permutation equivariance cannot be leveraged in ACE.
YE3T (bottom) is adapted to both rotations and permutations, and YE3T tensor products may be used for rotation/permutation equivariance because all physical-orbit information is retained until final coupling.}
\label{fig:sources}
\end{figure*}

As illustrated in Fig.~\ref{fig:sources}, the fixed-lattice cluster expansion only does symmetry adaptation (averaging over all point group symmetries of the parent lattice) in one final step to produce the ``orbit'' average \cite{sanchez_generalized_1984}.
Without adapting the fixed-lattice tensor product cluster functions to the lattice symmetry, there would be many times more tensor product functions to evaluate.
ACE symmetry adaptation (middle row in Fig.~\ref{fig:sources}) typically focuses on rotation symmetries. The reduced tensor products used in MLIPs and message-passing models can implicitly contain mixed $L>0$ information through ``intermediate'' states even when the output rotational symmetry type is $L=0$. However, ACE tensor products and the ACE basis are restricted to permutation-invariant outputs and permutation-invariant intermediates \cite{drautz_atomic_2019,drautz_atomic_2020,dusson_atomic_2022}.
There is no internal permutation information because it is averaged out early. Furthermore, speed and theoretical advantages gained from explicitly treating the true joint rotation/permutation character are lost.
In contrast, YE3T applies joint symmetry adaptation for rotations and permutations.
Rotation and permutation symmetry adaptation are handled simultaneously, and YE3T symmetry-adapted tensor products may implicitly contain both rotation and permutation information through intermediate states.
This is the benefit of the joint representation decomposition that YE3T provides, and it makes specific cases (like the permutation-invariant and rotation-equivariant basis construction) better-defined and faster than current approaches.

An important subset of these joint tensor product representations is the permutation-invariant (trivial permutation representation) sector.
The initially reported scalar and tensor ACE bases are obtained as one part of the full YE3T decomposition where the rotation and permutation content is trivial \cite{drautz_atomic_2019,dusson_atomic_2022,goff_permutation-adapted_2024}.
Tensor-valued ACE is the part of the YE3T decomposition with trivial permutation content and nontrivial rotation representation content \cite{drautz_atomic_2020}.
The same fully symmetric rotation carriers occur in equivariant message-passing methods such as NequIP, MACE, and GRACE \cite{batzner_e3-equivariant_2022,batatia_mace_2022,bochkarev_graph_2024}.

Beyond interatomic potentials and atomistic/electronic models, this work is relevant for other foundational methodologies and theories in the field. These include density-correlation methods and recursive relationships for higher-order permutation symmetry features \cite{nigam_recursive_2020,dusson_atomic_2022}.
While we do not discuss the conditioning and self-interactions present in density-projected features, they remain relevant.
In the Supplemental Material, we show nonlinear relationships between tensor product features of different ranks through primitive and decomposable subspaces.
The latter is a subspace of the full tensor product space whose tensor products can be composed from products of multiple lower-rank products.
Building our bases at the representation-theoretical level allows us to use computationally efficient density-projected features or self-interaction-free cluster interactions \cite{ho_atomic_2024}.

Independent Lie-algebra kernel methods construct rotation-equivariant, permutation-invariant spaces without specifying Clebsch--Gordan coupling paths \cite{barthelemy_efficient_2026} and provide a complete permutation-invariant, rotation-equivariant basis.
General Lie-group frameworks construct equivariant maps for broader carrier choices \cite{batatia_general_2023}.
The GE-PI construction in Barthelemy et al.~\cite{barthelemy_efficient_2026} provides a complementary construction of the fully symmetric parent $\lambda=(N)$, which is part of what is derived in this work; however, it extends to other Lie groups rather than to other permutation carriers.
Separate work accelerates complete Clebsch--Gordan tensor products \cite{xie_asymptotically_2026}. This is an implementation problem distinct from resolving fixed-content multiplicity spaces.
Nonetheless, we show that YE3T enables symmetry-adapted coupling constructions that reduce the computational cost of existing ACE models by a factor of $2$--$5$ with preliminary code and provide speedups of up to $1000\times$ relative to cuEquivariance in the reported Clebsch--Gordan coupling-construction benchmarks.
Overall, the representation-theoretical treatment presented here is immediately relevant to tensor products in ACE and methods with the same underlying rotation-carrier spaces, including NequIP, MACE, GRACE, e3nn modules, and SE(3)-equivariant attention \cite{batzner_e3-equivariant_2022,batatia_mace_2022,bochkarev_graph_2024,geiger_e3nn_2022,fuchs_se3-transformers_2020}.
The irreducible construction removes spurious redundancies reported in other works when applied to rotation-equivariant and permutation-invariant carrier spaces such as ACE.
It allows for an irreducible decomposition of joint rotation/permutation carriers and definitions of rotation/permutation-equivariant bases. Beyond ACE, it allows construction of tensor product bases with specific permutation character $\lambda$ and rotation character $L$, making it a natural extension to other important physics problems like electronic operator problems and fermionic functions (such as those in nuclear physics).

In this work, we derive the YE3T decomposition and use it to define general cluster/tensor product bases with arbitrary joint rotation/permutation character.
This is a more natural extension of the traditional cluster expansion, as we rely on the orbit of all joint physical symmetries instead of just rotations as ACE does.
We show that the ACE basis is a subset of the YE3T basis and that one may use the YE3T decomposition to obtain ACE basis functions and features more quickly than conventional methods (while still matching basis sizes of trusted methods).
Under the same representation theory, methods for nontrivial permutation targets are demonstrated in electronic-structure modeling.
In the following section, we summarize the mathematical derivation of the full YE3T basis and make detailed comparisons with previous work on ACE-like basis constructions. In particular, we confirm that YE3T reproduces all previously reported rigorous results for basis multiplicities, confirming the fundamental correctness of this new approach.
In Section~\ref{sec:models}, we outline several models that we have constructed using YE3T, including a simple linear ACE-like MLIP, a more complex message-passing MLIP, and a fermionic model for electron repulsion that incorporates nontrivial permutation symmetry.
In Section~\ref{sec:results}, we present preliminary results for the linear YE3T MLIP and the fermionic model.

\section{Theory}
\subsection{General tensor product (cluster) functions}
\label{sec:main_sources}
The YE3T representation theory and basis were developed to enable modeling of atomistic and electronic systems.
YE3T model construction is divided into three layers. The first concerns the cluster-interaction functions and which interactions remain explicit at each stage (e.g., before symmetry adaptation).
The representation theory for YE3T is used to decompose the carriers (the representation spaces) of joint $S_N\times SO(3)$ representations \cite[Chap.~1, Sec.~1]{serre_linear_1996}.
The main-text derivation is stated for $SO(3)$; the parity-aware $O(3)$ extension is given in the Supplemental Material.
The physical basis functions and descriptors are constructed by evaluating cluster functions before symmetry adaptation, then applying the coupling coefficients obtained from the carrier decomposition to adapt the basis to the appropriate joint rotation/permutation symmetries.
This three-part construction may begin with the single-interaction basis functions and the tensor product ``cluster'' interactions themselves, analogous to how the theory for traditional fixed-lattice cluster expansions is outlined.

The single-interaction basis, the $\phi$ basis from Drautz 2019 and Drautz 2020, may be defined as \cite{drautz_atomic_2019,drautz_atomic_2020}
\begin{equation}
\begin{aligned}
\phi_{ij\eta l m}(\boldsymbol r_{ij},\mu_i,\mu_j,x_{ij},\ldots)
&=
\left[R_{nl}^{\mu_i\mu_j}(r_{ij})T_\omega(x_{ij})\cdots\right]
Y_l^m(\widehat{\boldsymbol r}_{ij}),
\end{aligned}
\label{eq:main_phi_basis}
\end{equation}
where $r_{ij}$ is the radial distance between central and neighboring particles, $\mu_i$ and $\mu_j$ are their chemical labels, $R_{nl}^{\mu_i\mu_j}$ is a species-dependent radial basis, and $Y_l^m$ is a spherical harmonic basis for spatial angular degrees of freedom between central and neighboring particles.
The variable $x_{ij}$ and the basis function $T_\omega(x_{ij})$ account for additional non-angular degrees of freedom and may include continuous scalar properties such as charge.
All non-angular degrees of freedom are grouped into a collected index, $\eta=(\mu_i,\mu_j,n,\omega,\ldots)$, and this is combined with the angular momentum index $l$ to form the overall basis index $\nu=(\eta,l)$.
For the YE3T decomposition, the repeated channel content relevant for Schur--Weyl decomposition later is classified by repeated $\nu=(\eta,l)$.

In practice, the values of the $\phi$ basis are often summed over neighbors to form the density-projected functions, and these have the same underlying rotational carriers.
\begin{equation}
A_{i\nu m}
=
\sum_{j\ne i}\phi_{ij\nu m}(\boldsymbol r_{ij},\mu_i,\mu_j,x_{ij},\ldots).
\label{eq:main_density_basis}
\end{equation}
This ``density-projected basis'' is formed by summing over all neighbors $j$ within some physical cutoff.
The result is that $A_{i\nu m}$ is symmetric with respect to the exchange of physical neighbors.
As a result, their tensor products can only occupy the fully symmetric $\lambda=(N)$ symmetry type \cite{drautz_atomic_2019,dusson_atomic_2022}.

There are other ``single-interaction'' functions that may be used in place of the explicit $\phi$ basis or the density-projected basis $A$.
As long as it transforms in the rotation carrier (and later in the permutation carrier) correctly, it may be used for YE3T tensor product constructions.
Thus, single-interaction basis functions will generally be denoted as $u_{i,f;\nu m}$ where $f$ is an index denoting the factor position in a tensor product cluster function.
A full $N$-factor tensor product may be written as a product of generic single-interaction functions that transform in the angular carrier space for $\nu=(\eta,l)$.
\begin{equation}
Z_{i;\boldsymbol\nu^\circ,\mathbf m}
=
\prod_{f=1}^{N}
u_{i,f;\nu_f^\circ m_f},
\qquad
\boldsymbol\nu^\circ=(\nu_1^\circ,\ldots,\nu_N^\circ).
\label{eq:main_source_product}
\end{equation}
Here $\mathbf m=(m_1,\ldots,m_N)$ and a canonically ordered tensor factor is denoted as $\boldsymbol{\nu}^\circ=(\nu^\circ_1,\nu^\circ_2,\ldots,\nu^\circ_N)$. This canonical ordering will be defined later, but in general, any other ordering of the tensor product may be written in terms of permutations of a canonical reference ordering $\boldsymbol{\nu}^\circ$. These permutations will be denoted as $\boldsymbol{\mu} = \sigma \cdot \boldsymbol{\nu}^\circ$.
For the functions in Eq.~\eqref{eq:main_source_product}, the factor ordering matters.

The typical ACE construction uses $u_{i,f;\nu_f^\circ m_f}=A_{i\nu_f^\circ m_f}$ for each factor in a tensor product. Explicit $\phi$ constructions use $\phi_{ij,\nu m}$ for each factor. Other single-interaction sources discussed in the Supplemental Material include a tagged-neighbor basis that combines explicit $\phi_{ij \, \nu m}$ with neighbor-summed coordinates and lifted density projections (which are effectively density-projected functions, but with shells around a central atom).
However, the representation theory may be extended to various versions of them, so we will generally talk about the generic $u_{i,f;\nu_f^\circ m_f}$ basis, reminding the reader that it may be applied to various single-interaction sources.

\subsection{Permutation structure in ACE-like tensor products}

The previous subsection discussed single-interaction functions that transform in the rotation carriers.
The rotation carriers for a given single-interaction function are the rotation carrier spaces corresponding to the spherical harmonics with clear quantum mechanical analogs \cite[Chap.~3, Sec.~3]{SakuraiNapolitano2020}.
Here, we also refer to the rotational carrier space as $V_l$; however, we need to describe the carrier of the overall tensor product and how it reduces.
In general, a tensor product of rotational representation carriers is reducible \cite{yutsis_mathematical_1962}.
To derive the YE3T representation decomposition, we define the carrier for the joint rotation/permutation representations that underpin ACE-like tensor products.
This is the carrier space of $S_N{\times}SO(3)$ representations \cite{bacon_efficient_2006}.
As a tensor product space, it is composed of one-particle channel spaces that individually carry irreducible representations of $SO(3)$, denoted by $V_{l_f}$ or, if non-angular channels produce different copies of an angular carrier space, $\mathcal V_{\nu_f}$.
In both cases, the index $f$ labels a factor in the tensor product.

The fixed channel content is denoted by $\boldsymbol\nu$, with $\boldsymbol\nu^\circ$ the ordered reference tuple introduced in Eq.~\eqref{eq:main_source_product}.
A generic ordering is denoted by $\boldsymbol\mu=\sigma\!\cdot\!\boldsymbol\nu^\circ$, generated by some permutation $\sigma\in S_N$.
A specific canonical reference ordering is often chosen for convenience or to avoid obvious repetitions when enumerating functions.
Lexical ordering of tensor product indices is common \cite{drautz_atomic_2019,dusson_atomic_2022}, but any canonical ordering scheme can be used, such as frequency-based ordering of indices \cite{goff_permutation-adapted_2024}.
This frequency-based ordering makes the stabilizing permutation group more obvious.
We adopt a frequency-based ordering as the canonical tensor product ordering for YE3T, but this is an arbitrary choice.
In this canonical ordering scheme, distinct channel values $\nu_f$ are ordered first by frequency $k_b$ and second by the lexicographic ordering of channel tuples $\nu_b$.
When using the frequency-based reference ordering, this stabilizer is the Young subgroup $G_{\boldsymbol\nu}=\prod_{b=1}^{B}S_{k_b}$, where repeated channel indices are grouped into blocks of size $k_b$ and $N=\sum_b k_b$.
The distinct orderings generated by the reference tuple are contained in the set $\operatorname{Orb}(\boldsymbol\nu^\circ)$.
As a concrete example, for $\boldsymbol\nu^\circ=(B,B,A)$, $\operatorname{Orb}(\boldsymbol\nu^\circ)=\{(B,B,A),(B,A,B),(A,B,B)\}$.
\begin{equation}
\left|\operatorname{Orb}(\boldsymbol\nu^\circ)\right|
=
\frac{|S_N|}{|\operatorname{Stab}(\boldsymbol\nu^\circ)|}
=
\frac{N!}{\prod_{b=1}^{B}k_b!}.
\label{eq:main_orbit_stabilizer}
\end{equation}
Equation~\eqref{eq:main_orbit_stabilizer} is the orbit--stabilizer relation for the action of the full group $S_N$ on ordered channel tuples, with stabilizer given by the corresponding Young subgroup \cite[Sec.~1.3]{james_representation_1984}.
The joint rotation/permutation tensor-product carrier is defined by taking this orbit stabilization into account:
\begin{equation}
\widehat{\mathcal H}_{\boldsymbol\nu}
=
\bigoplus_{\boldsymbol\mu\in\operatorname{Orb}_{S_N}(\boldsymbol\nu^\circ)}
\mathcal H_{\boldsymbol\mu},
\qquad
\mathcal H_{\boldsymbol\mu}
=
\bigotimes_{f=1}^{N}\mathcal V_{\mu_f}.
\label{eq:main_orbit_closed_carrier}
\end{equation}
The expansion includes all distinct permutations of the reference ordering without repeated patterns. For example, $\widehat{\mathcal H}_{\boldsymbol\nu}$ does not contain repeated copies such as $[(B,B,A),(B,B,A),\ldots]$: although the transposition $(12)$ swaps positions 1 and 2, it leaves $\boldsymbol{\nu}=(B,B,A)$ unchanged. This is what is meant by distinct in this context.

The definition above is actually what makes $\widehat{\mathcal H}_{\boldsymbol\nu}$ the carrier for the joint $S_N\times SO(3)$ representations and not just $SO(3)$. Without it, a permutation representation of $S_N$ may not live in the carrier space.
Further consideration of Young-subgroup symmetry adaptation follows naturally from the Young-subgroup stabilizer. Rather than always building all $S_N$-equivariant mappings, we can systematically introduce permutation character with Young subgroups of $S_N$, up to the case where the Young subgroup $G\le S_N$.
Here, we exploit Young subgroups that lie within the stabilizing subgroup of repeated tensor product content, but this is not a requirement of the method.
Permutations and rotations commute.
\begin{equation}
P(\sigma)\lvert\boldsymbol\mu,\mathbf m\rangle
=
\lvert\sigma\!\cdot\!\boldsymbol\mu,
       \sigma\!\cdot\!\mathbf m\rangle,
\qquad
[P(\sigma),D(R)]=0,
\quad
(\sigma,R)\in S_N\times SO(3).
\label{eq:main_commuting_actions}
\end{equation}
Again, we generally discuss these permutation actions as permutations of tensor product factors. For "single-interaction" constructions in which physical interaction roles are tracked, they can be viewed as actions induced by physical-neighbor permutations.
This commutation allows us to perform the decomposition as we do.

\subsection{Schur--Weyl YE3T decomposition}

The blockwise Schur--Weyl decomposition that leads to the full decomposition of Eq.~\eqref{eq:main_orbit_closed_carrier} is straightforward.
Schur--Weyl decomposition has been suggested to treat the joint rotation/permutation representation content in ACE-like tensor products before \cite{dusson_atomic_2022}.
However, Schur--Weyl decomposition is not applied to carriers with mixed $SO(3)$ representations \cite[Sec.~6.1]{fulton_representation_2013}\cite{bacon_efficient_2006}.
The approach taken here does not apply Schur--Weyl decomposition to the full tensor-product space, but applies it to ``blocks'' with repeated channel content where the rotational carrier is shared within the block.
A block in this context is defined as a subset of factor indices that share the same $\nu,l$, and is the same definition provided in Eq. \eqref{eq:main_orbit_stabilizer}.

The key step in the YE3T decomposition of the $S_N\times SO(3)$ carrier is applying Schur--Weyl decomposition within each block and restricting the resulting general-linear carrier to $SO(3)$.
This restriction is required because Schur--Weyl typically applies to general linear groups, not just $SO(3)$ \cite[Sec.~6.1]{fulton_representation_2013}.

\begin{figure*}[t]
\color{black}
\centering
\includegraphics[width=0.98\textwidth]{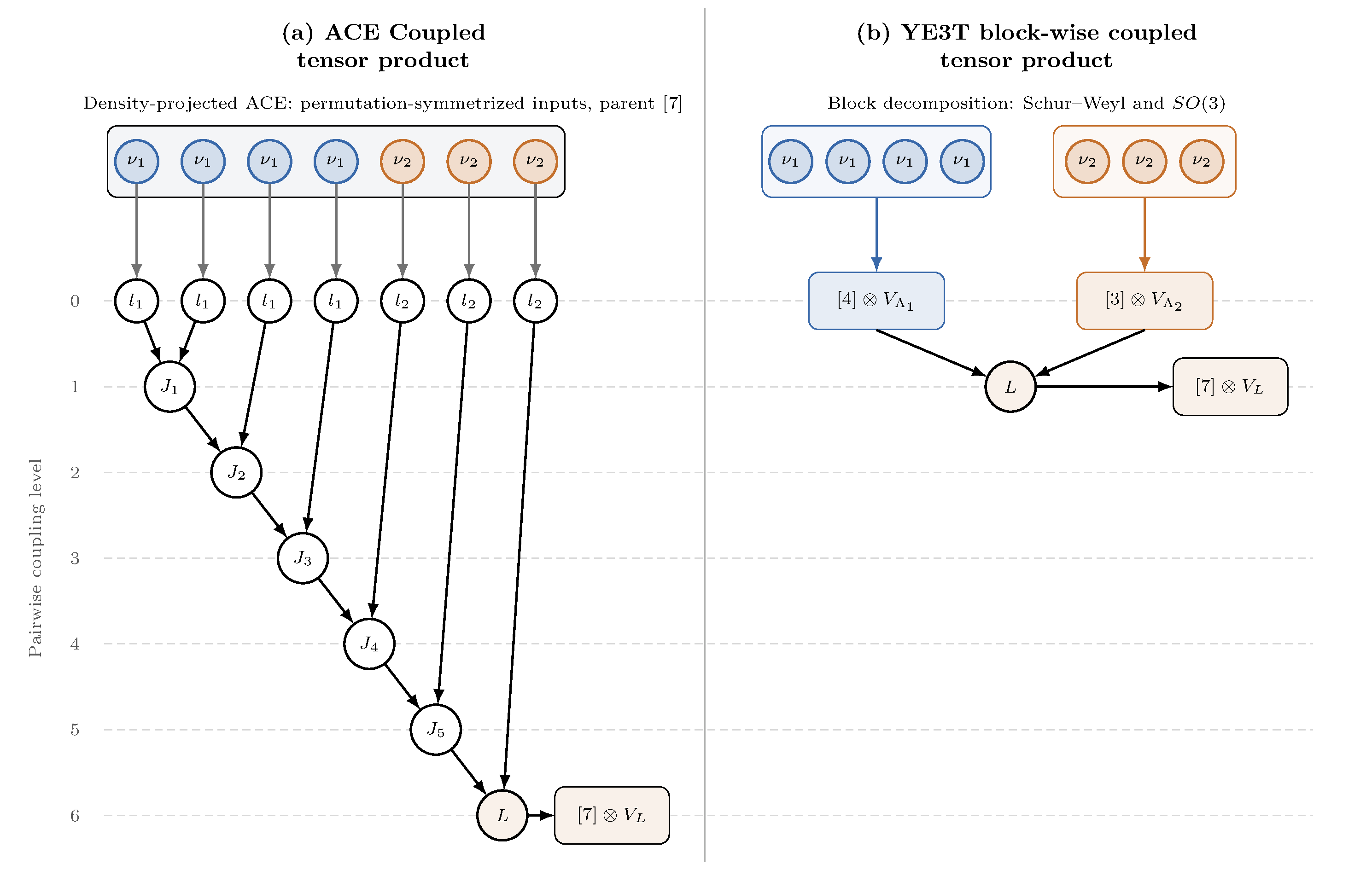}
\caption{Comparative illustrations of the ACE and YE3T coupled tensor product constructions for the same tensor product, $\boldsymbol\nu^\circ=(\nu_1,\nu_1,\nu_1,\nu_1,\nu_2,\nu_2,\nu_2)$, with the permutation-invariant parent type $\lambda=(7)$. (a) The density-projected ACE construction begins from permutation-symmetrized input factors and uses six binary coupling levels \cite{drautz_atomic_2019,dusson_atomic_2022}. (b) The YE3T coupling construction builds the tensor product through one intermediate level following blockwise Schur--Weyl decomposition.}
\label{fig:main_cg_vs_ye3t}
\end{figure*}

The decomposition for each block of size $k_b$ is given by
\begin{equation}
\mathcal V_{\nu_b}^{\otimes k_b}
\cong
\bigoplus_{\kappa_b\vdash k_b}\bigoplus_{\Lambda_b}
[\kappa_b]\otimes V_{\Lambda_b}\otimes
\mathbb C^{d_b^{\kappa_b\Lambda_b}}.
\label{eq:main_block_decomposition}
\end{equation}
In Eq.~\eqref{eq:main_block_decomposition}, $\kappa_b$ is a representation of $S_{k_b}$, which may be written as a Young diagram corresponding to that permutation irreducible representation (and, in shorthand, as a partition of $k_b$).
The multiplicity space is $\mathbb{C}$ with dimension $d_b^{\kappa_b\Lambda_b}$.
This $d_b^{\kappa_b\Lambda_b}$ is the multiplicity of the irreducible $S_{k_b}\times SO(3)$ carrier $[\kappa_b]\otimes V_{\Lambda_b}$ within block $b$.
Further details on this decomposition and on restricting the general-linear carrier to $SO(3)$ may be found in the Supplemental Material.
This blockwise decomposition followed by recoupling of block-decomposed states is faster than angular coupling of every rotational state (as is done in ACE constructions).
This is illustrated in Fig.~\ref{fig:main_cg_vs_ye3t}, which compares a typical recursive ACE coupling with a balanced pairwise-YE3T coupling that has been block-symmetry adapted.
The shallower tensor product tree is reflected in computational efficiency as demonstrated later.

The complete tensor-product carrier in Eq.~\eqref{eq:main_orbit_closed_carrier} can be composed of multiple blocks of repeated tensor-product content, and Eq.~\eqref{eq:main_block_decomposition} must be repeated for each block.
The resulting block carriers must be recoupled to the final parent states.
This is done by taking tensor products between the block intermediates, which are then coupled to the final parent carrier.
After the blockwise Schur--Weyl decomposition, the intermediate block-to-parent coupling for rotation carriers can be treated simultaneously with the intermediate block-to-parent coupling for permutation carriers because the two actions commute.
\begin{subequations}
\label{eq:main_block_to_parent_carriers}
\begin{align}
[\boldsymbol\kappa]\uparrow_{G_{\boldsymbol\nu}}^{S_N}
&\cong
\bigoplus_{\lambda\vdash N}
\mathbb C^{c_{\boldsymbol\kappa}^{\lambda}}\otimes[\lambda],
\label{eq:main_subgroup_induction_arrows}
\\[3pt]
\bigotimes_{b=1}^{B}V_{\Lambda_b}
&\cong
\bigoplus_L
\mathcal M_{\boldsymbol\Lambda}^{L}\otimes V_L,
\qquad
M_{\boldsymbol\Lambda}^{L}:=\dim\mathcal M_{\boldsymbol\Lambda}^{L}.
\label{eq:main_prodgroup_restriction_arrows}
\end{align}
\end{subequations}
The equivalent subduction expression is given in the Supplemental Material.
The block rotation carriers are handled with typical Clebsch--Gordan couplings \cite[Chs.~I--II]{yutsis_mathematical_1962}.
In Eq.~\eqref{eq:main_block_to_parent_carriers}, we have shown the induction from the Young product carrier to the symmetric group $S_N$ using the James--Kerber notation \cite{james_representation_1984}.
The product carrier, $[\boldsymbol\kappa] =\boxtimes_{b=1}^{B}[\kappa_b]$, is formed by taking the external tensor product of the block carriers.

The equivalent subduction expression for Eq.~\eqref{eq:main_subgroup_induction_arrows} is given in the Supplemental Material. In practice, we may use the subduction coefficients produced by restricting from $S_N$ to $G_{\boldsymbol{\nu}^\circ}$ to obtain the parent permutation carrier from the product permutation carriers \cite[Secs.~1.3, 2.2, and 2.8]{james_representation_1984}.
Block rotation carriers are handled with typical Clebsch--Gordan couplings \cite[Chs.~I--II]{yutsis_mathematical_1962}.
These equations describe the recoupling of block rotation/permutation carriers to parents.
Different complete coupling trees amount to different choices of basis within the same multiplicity space, related by unitary transformations \cite[Chs.~II--VI]{yutsis_mathematical_1962}.
The YE3T decomposition is not limited to a particular coupling tree or formalism; however, we rely more on balanced pairwise schemes and graphs because they result in shallower binary trees; cf.~Fig.~\ref{fig:main_cg_vs_ye3t}.

The multiplicities $c_{\boldsymbol\kappa}^{\lambda}$ denote how many copies of the parent carrier $[\lambda]$ occur in the representation induced from the Young-subgroup carrier $[\boldsymbol\kappa]$ to $S_N$.
The same integer, by Frobenius reciprocity, counts copies of $[\boldsymbol\kappa]$ when $[\lambda]$ is restricted to the Young subgroup and is a Littlewood--Richardson multiplicity \cite[Secs.~1.3, 2.2, and 2.8]{james_representation_1984}.
The representation-theoretical decomposition in YE3T follows from the decomposition of Eq.~\eqref{eq:main_orbit_closed_carrier} into its irreducible components:
\begin{equation}
\widehat{\mathcal H}_{\boldsymbol\nu}
\cong
\bigoplus_{\lambda\vdash N}\bigoplus_L
[\lambda]\otimes V_L\otimes
\mathscr M_{\boldsymbol\nu}^{\lambda L}.
\label{eq:main_ye3t_decomposition}
\end{equation}
In Eq.~\eqref{eq:main_ye3t_decomposition}, the permutation carrier is denoted by $[\lambda]$, $V_L$ is the parent angular carrier, and $\mathscr M_{\boldsymbol\nu}^{\lambda L}$ is the multiplicity space for the full joint decomposition. The block and parent multiplicity spaces combine multiplicatively through a tensor product to form this full multiplicity space.
The multiplicity space defines explicit rotation/permutation-equivariant basis counts as well as the multiplicity coordinates that yield a complete, independent basis for a rank-$N$ tensor-product space with fixed channel content $\boldsymbol\nu$.

Summing the corresponding products of multiplicities over allowed block labels $(\boldsymbol\kappa,\boldsymbol\Lambda)$ that couple to the output $(\lambda,L)$ gives the irreducible multiplicity
\begin{equation}
\begin{aligned}
\alpha_{\boldsymbol\nu}^{\lambda L}
&=\dim\mathscr M_{\boldsymbol\nu}^{\lambda L}
=
\sum_{\boldsymbol\kappa}
\sum_{\boldsymbol\Lambda}
c_{\boldsymbol\kappa}^{\lambda}
d_{\boldsymbol\kappa\boldsymbol\Lambda}
M_{\boldsymbol\Lambda}^{L},
\\
d_{\boldsymbol\kappa\boldsymbol\Lambda}
&=\prod_{b=1}^{B}d_b^{\kappa_b\Lambda_b}.
\end{aligned}
\label{eq:main_multiplicity_count}
\end{equation}
In Eq.~\eqref{eq:main_multiplicity_count}, $d_{\boldsymbol\kappa\boldsymbol\Lambda}$ is the product of the blockwise Schur--Weyl/rotational-branching multiplicities from Eq.~\eqref{eq:main_block_decomposition}.
The permutation coupling multiplicity $c_{\boldsymbol\kappa}^{\lambda}$ is the Littlewood--Richardson induction/subduction multiplicity \cite[Secs.~1.3, 2.2, and 2.8]{james_representation_1984}.
The angular multiplicity $M_{\boldsymbol\Lambda}^{L}$ is the number of independent angular-momentum recoupling paths \cite[Chs.~II--VI]{yutsis_mathematical_1962}.
This formula provides an analytical count of irreducible joint rotation/permutation representations with parent Young symmetry $\lambda$ and angular momentum $L$.
We use this to evaluate exactly the number of independent basis functions for high-rank tensor products without approximation (up to $N=16384$), as reported in the Supplemental Material.

The full decomposition produces an extension of the typical $(2L+1)$ dimension identity of a decomposed $SO(3)$ carrier, $V_L$. Accounting for the joint parent permutation representation space, we obtain the dimension relationship
\begin{equation}
\frac{N!}{\prod_bk_b!}
\prod_{b=1}^{B}\left(\dim\mathcal V_{\nu_b}\right)^{k_b}
=
\sum_{\lambda\vdash N}\sum_L
\dim[\lambda]\,(2L+1)\,
\alpha_{\boldsymbol\nu}^{\lambda L}.
\label{eq:main_dimension_identity}
\end{equation}
Here, the dimension of $[\lambda]$ is the number of standard Young tableaux of shape $\lambda$ \cite[Sec.~3.4]{james_representation_1984}. This is the permutation analog of the angular projection quantum numbers $M$ that build up the $(2L+1)$ size of the rotation carrier $V_L$.

\subsection{The YE3T basis, equivariance, orthonormality, and completeness}

The permutation and rotation actions commute. Thus, the rotation and permutation couplings in Eq.~\eqref{eq:main_block_to_parent_carriers} combine multiplicatively, and joint permutation--rotation coupling coefficients can be defined.
An orthonormal basis follows naturally from the carrier decomposition.
We use a compact label for the uncoupled tensor-product indices $q=(\boldsymbol\mu,\mathbf m)$, with $\mathbf m=(m_1,\ldots,m_N)$.
A coupled output basis label is indexed by $\upalpha=(\lambda,L,a,t,M)$, where $M$ indexes the magnetic component of $V_L$, $t$ indexes a component of the parent Young carrier $[\lambda]$, and $a=1,\ldots,\alpha_{\boldsymbol\nu}^{\lambda L}$ indexes the irreducible multiplicity copy.
Let $|q\rangle$ and $|\upalpha\rangle$ denote the corresponding uncoupled and coupled basis vectors, respectively.

Before coupling, the factors in the tensor product basis are denoted by $u_{i,f;\mu_fm_f}$.
The corresponding raw feature vector is $|Z_i\rangle=\sum_q Z_{i,q}|q\rangle$, following the $Z=\prod u_f$ notation defined earlier. The coupled basis is denoted as $B_{i,\upalpha}$.
We define the joint rotation/permutation coupling coefficients by
\begin{subequations}
\label{eq:main_ye3t_coupling}
\begin{align}
\mathsf C_{q\upalpha}
&:=
\langle\upalpha|q\rangle,
\\
|\upalpha\rangle
&=
\sum_q\mathsf C_{q\upalpha}^{*}|q\rangle,
\label{eq:main_ye3t_coupled_basis}
\\
B_{i,\upalpha}
&=
\langle\upalpha|Z_i\rangle
=
\sum_q\mathsf C_{q\upalpha}Z_{i,q},
\label{eq:main_ye3t_descriptor}
\\
\sum_q\mathsf C_{q\upalpha}\mathsf C_{q\upbeta}^{*}
&=\delta_{\upalpha\upbeta},
\qquad
\sum_{\upalpha}\mathsf C_{q\upalpha}^{*}\mathsf C_{q'\upalpha}
=\delta_{qq'}.
\label{eq:main_ye3t_orthogonality_completeness}
\end{align}
\end{subequations}
The sum over $q$ runs over the distinct orderings $\boldsymbol\mu\in\operatorname{Orb}_{S_N}(\boldsymbol\nu^\circ)$ and their allowed magnetic components $\mathbf m$.
The sum over $\upalpha$ in the second identity runs over the complete coupled basis, while $q'$ is a second uncoupled basis label and $\upbeta$ is a second coupled label of the same form as $\upalpha$.
The first identity in Eq.~\eqref{eq:main_ye3t_orthogonality_completeness} states orthonormality of the coupled basis; the second states completeness on the full orbit-closed carrier.
The joint coupling coefficients $\mathsf C$ are analogues of generalized Clebsch--Gordan coefficients for purely angular tensor products \cite[Chs.~II--VI]{yutsis_mathematical_1962}.
These coefficients construct the coupled basis vectors in Eq.~\eqref{eq:main_ye3t_coupled_basis} and transform the uncoupled tensor products in Eq.~\eqref{eq:main_source_product} into the coupled coordinates in Eq.~\eqref{eq:main_ye3t_descriptor}.
The orthonormality and completeness relations are discussed in more detail in the Supplemental Material.
The joint coupling coefficients, including Young--Yamanouchi/Young-orthogonal and subduction/induction conventions, are also derived there \cite[Secs.~2.2, 2.8, and 3.4]{james_representation_1984}\cite{chilla_linear_2006,chilla_reduced_2006}.

Selecting the fully symmetric parent type $\lambda=(N)$ gives the ACE sector of the YE3T basis.
For $L=0$, this gives the scalar ACE basis in Ref. \citenum{drautz_atomic_2019}; nonzero $L$ gives tensor-valued, permutation-invariant components.
However, YE3T is not limited to trivial permutation representations.
Specific $(\lambda,L)$ pairs may be retained.

\section{Models and Selected Applications}
\label{sec:models}
\subsection{Physical Descriptors and Linear Expansions}
The physical descriptor evaluation depends on the model type (e.g., what particles and what cluster-interaction types are being evaluated).
For example, if the factor functions are density basis functions, they may be evaluated based on physical neighbor-list projection of $\phi$, as is done in Ref.~\citenum{drautz_atomic_2019}.
If the factor functions are explicit $\phi$ interactions, or a mixed explicit-factor/density (a.k.a.\ tagged) function, then neighbor indices are summed only at the correct pooling stage. For an explicit cluster interaction built from $\phi$ functions, this corresponds to an explicit cluster motif $\mathsf M$, and the sum is over allowed placements of the physical cluster.

The simplest YE3T model one can produce is a tensor product basis expansion.
This linear expansion is analogous to those used for the fixed-lattice cluster expansion and ACE.
One may expand a scalar property, like the energy for an MLIP, as a function of YE3T features as
\begin{equation}
T=T_{ref}+\sum_i\sum_\upalpha c_\upalpha B_{i\upalpha},
\label{eq:main_fixed_descriptor_energy}
\end{equation}
For an MLIP, the property $T$ in Eq.~\eqref{eq:main_fixed_descriptor_energy} is the energy $E$, and the reference value $T_{ref}$ is the isolated-atom energy, scaled to the system size because it lies outside the sum over atoms $i$. The expansion may also be used for other extensive quantities and for non-atomistic systems.
The coefficients $c_\upalpha$ are fitted linear expansion coefficients and $\upalpha$ is a combined index that carries all information needed for an unambiguous basis label, $(\lambda,L,a,t,M)$, which includes the overall joint parent symmetry type $(\lambda,L)$, the multiplicity index $a$, and the permutation and rotation component indices $t$ and $M$, respectively.
Through a Cauchy basis definition, one may have linear models and descriptors, $\mathcal{B}_{i\upalpha}$ with non-trivial intermediate rotation and permutation states. In other words, the intermediate coupled permutation states do not always need to be permutation invariant to get a permutation invariant feature/output.

\subsection{Rotation/permutation Equivariant Tensor Message Passing Models}

YE3T message-passing follows naturally from the joint rotation/permutation structure developed in earlier sections.
For an input construction $F$ of a cluster/tensor input before joint symmetry adaptation, we denote its uncoupled coordinates by $Z_{iF,q}^{(\tau)}$ for the root particle/atom $i$ and layer $\tau$.
Consistent with Eq.~\eqref{eq:main_ye3t_coupling}, the joint rotation/permutation coupling coefficients for this construction are denoted by $\mathsf C^{(F)}_{q\upalpha}=\langle\upalpha|q\rangle$.
For each coupling, we obtain
\begin{equation}
B_{iF,\upalpha}^{(\tau)}
=
\sum_q \mathsf C^{(F)}_{q\upalpha}Z_{iF,q}^{(\tau)}.
\label{eq:main_message_source_projection}
\end{equation}
The resulting $N$-factor components from Eq.~\eqref{eq:main_message_source_projection} transform in $[\lambda]\otimes V_L$ of $S_N\times SO(3)$ (or $O(3)$ if using the $O(3)$ extension outlined in the supplemental).
Recall that $\lambda\vdash N$, where $\lambda$ labels the tensor-factor permutation symmetry and $L$ labels the rotation symmetry.
Nontrivial Young components can be propagated when supported by the
input construction.

Families of input constructions are left intentionally general. We do not restrict them to density products.
We denote the input-construction family by $\mathcal X_i^{(\tau)}$; it may combine density projections, explicit cluster motifs/cluster functions $\mathsf{M}$, tagged cluster functions, or other functions that transform correctly in the $[\lambda]\otimes V_L$ carrier space.
This connects to existing graph-based and density-based message-passing models, including methods with single distinguishable input sources, but is not restricted to such constructions \cite{batatia_mace_2022,bochkarev_graph_2024,musaelian2023learning,nigam2022equivariant}.

In standard message-update notation,
equivariant message and update maps $\mathcal M_F^{(\tau)}$ and
$\mathcal U^{(\tau)}$ give
\begin{align}
m_i^{(\tau)}
&=
\operatorname*{AGG}^{(\tau)}_{F\in\mathcal X_i^{(\tau)}}
\mathcal M_F^{(\tau)}\!\left(h_i^{(\tau)},B_{iF}^{(\tau)}\right),
\label{eq:main_message_carrier_aggregation}
\\
h_i^{(\tau+1)}
&=
\mathcal U^{(\tau)}\!\left(h_i^{(\tau)},m_i^{(\tau)}\right),
\label{eq:main_message_update}
\end{align}
where $h_i^{(\tau)}$ collects the hidden features, with any retained
neighbor or cluster indices implicit \cite{gilmer_neural_2017}.
The aggregation $\mathrm{AGG}$ preserves the joint symmetry, with contributions summed after alignment of joint representations and channel bases.
Learned linear maps mix channels and representation copies.
The product couplings allow different symmetry types to interact using fixed-rank or rank-increasing rotation/permutation couplings.
Message and update maps may combine these products with residual connections, scalar gates, attention with invariant weights, or other nonlinearities that respect the joint equivariance \cite{fuchs_se3-transformers_2020}.

This approach allows for different binary coupling trees and interaction graphs; they are independent.
For a $T$-layer model, a readout $\mathcal R$ with targeted rotation/permutation symmetry (e.g., rotation and permutation invariance) may use one or several
hidden states to give the machine-learned property.
For example, atomic energy in a machine-learned potential may be expressed as
\begin{equation}
E_i=\mathcal R\!\left(h_i^{(0)},\ldots,h_i^{(T)}\right).
\label{eq:main_message_readout}
\end{equation}
Overall, YE3T specifies the joint representation and coupling structure, while input features, aggregation, nonlinear updates, and coupling trees remain architectural choices.

\subsection{Electronic-repulsion-integral model}

To demonstrate another target with a different joint rotation/permutation symmetry type, we learn the electron--electron repulsion interaction in an antisymmetric orbital-pair basis.
For a given molecular geometry $X$, we will denote a L\"owdin-orthogonalized orbital as $|a\rangle$ \cite{lowdin_nonorthogonality_1950}.
The electron--electron repulsive Coulomb-interaction operator in atomic units is $\hat v_{12}=r_{12}^{-1}$, giving the geometry-dependent integrals in chemists' notation as
\begin{equation}
    (ab|cd)_X=\langle ac|\hat v_{12}|bd\rangle_X.
    \label{eq:ee_integral}
\end{equation}
We use normalized wedge states
\[
|a\wedge b\rangle
=
\frac{|ab\rangle-|ba\rangle}{\sqrt{2}},
\qquad a<b,
\]
which form an orthonormal basis of the antisymmetric two-electron space
\cite[Secs.~5.3--5.4]{derezinski_representations_2006}.
From this, we define our ERI matrix target:
\begin{equation}
\begin{aligned}
V_{ab,cd}(X)
&=\langle a\wedge b|\hat v_{12}|c\wedge d\rangle_X
\\
&=(ac|bd)_X-(ad|bc)_X.
\end{aligned}
\label{eq:ERI_target}
\end{equation}
We learn $X\mapsto\mathbf V(X)$ for
$\mathrm{H}_2\mathrm{CO}_3$ and $\mathrm{CO}_3^{2-}$ as example molecular systems.
The pairwise antisymmetrizer $\mathcal A_{ab}\mathcal A_{cd}=\tfrac14[1-(ab)][1-(cd)]$ selects the Young-subgroup type $[1,1]\boxtimes[1,1]$. Together with symmetry under pair exchange and the cyclic identity $V_{ab,cd}+V_{ac,db}+V_{ad,bc}=0$, which follows from Eq.~\eqref{eq:ERI_target}, this gives the parent permutation symmetry type $\lambda=(2,2)$ \cite{fiedler_structure_2003}.
The joint permutation--rotation-equivariant message passing described
above is applied to this wedge representation, rather than an ACE-like density basis.

To train the operator models, we generated 100 molecular-geometry configurations with mixed bond angles and distances for $H_2CO_3$ and $CO_3^{2-}$, for which proton angles and distances were also varied.
For these structures, we evaluated the ERI matrix entries using the cc-pVDZ basis through PySCF \cite{dunning_gaussian_1989,sun_pyscf_2018,li_introducing_2025}.
For this dataset, we trained two types of YE3T-equivariant message-passing models using site-basis functions for electronic interactions: the first retained mixed internal permutation character and mixed internal rotation character, while the second retained fixed pair-antisymmetric permutation character and mixed internal rotation character throughout the model.
The first type requires pair antisymmetry only at readout and allows the model to learn mixed joint rotation/permutation paths; the second is restricted to pair-antisymmetric representations throughout, including its hidden features.
The second is analogous to classical models restricted to permutation-invariant representations throughout.

\section{Results}
\label{sec:results}

The joint treatment of rotation/permutation content enables explicit multiplicity-resolved constructions of the ACE basis and more general YE3T bases with mixed permutation character.
There are no comparable references in the literature that enumerate basis functions and multiplicities for general $S_N\times SO(3)$ product representations to this extent.
However, we can validate against ACE basis sizes from trusted methods \cite{barthelemy_efficient_2026}.
This reduces to a comparison of the fully symmetric YE3T parent $\lambda=(N)$ across $L$ sectors, where $L$ may take any value allowed by the polygon conditions defined by Yutsis, Ref.~\citenum{yutsis_mathematical_1962}.
Barthelemy \textit{et al.} obtain the corresponding basis as a Lie-algebra kernel space, which they call the GE-PI basis \cite{barthelemy_efficient_2026}. YE3T instead obtains it using the YE3T representation decomposition.
In Table~\ref{tab:main_gepi_count_comparison}, we report the symmetric $\lambda=(N)$ basis counts for all allowed $L$.
For different ranks and $(\eta,l)$ combinations that may occur in an ACE model, the table reports the all-$L$ multiplicity $\sum_L\alpha_{\boldsymbol\nu}^{(N)L}$ predicted by YE3T and GE-PI, its distribution over $L$, and the per-$L$ contributions. The $L=0$ entries give the basis functions often used as descriptors in linear ACE models.
YE3T and GE-PI agree on all total and per-$L$ values.
The YE3T-symmetric and GE-PI approaches are not guaranteed to produce the same exact basis functions (they differ by coupling coefficients), but they are related by a unitary transformation.

\begin{table*}[t]
\color{black}
\centering
\scriptsize
\setlength{\tabcolsep}{3.5pt}
\renewcommand{\arraystretch}{1.08}
\caption{
Fully symmetric-parent, ACE-like $\lambda=(N)$ multiplicities for selected combinations of tensor product channel content.
The two middle columns report the total multiplicity $\sum_L\alpha_{\boldsymbol\nu}^{(N)L}$: the YE3T column gives the number produced by the representation-theoretical decomposition, and the GE-PI column gives the number produced by the nullspace-kernel method of Barthelemy et al. The final column gives the nonzero multiplicity at each $L$. The final row reproduces the symmetric parent in Table~\ref{tab:main_rank6_parent_multiplicities}.}
\label{tab:main_gepi_count_comparison}
\begingroup\color{black}
\newcommand{\countrowsep}{\arrayrulecolor{black!25}\specialrule{0.2pt}{1.2pt}{1.2pt}\arrayrulecolor{black}}
\begin{tabular}{c c c r r l}
\toprule
$N$ & $\boldsymbol \eta$ & $\boldsymbol l$
& \multicolumn{2}{c}{$\displaystyle\sum_L\alpha_{\boldsymbol\nu}^{(N)L}$}
& $L{:}\alpha_{\boldsymbol\nu}^{(N)L}$ \\
\cmidrule(lr){4-5}
& & & YE3T & GE-PI & \\
\midrule
2 & $(1,1)$ & $(1,1)$
  & 2 & 2 & $0{:}1,\ 2{:}1$ \\
\countrowsep
2 & $(1,2)$ & $(1,1)$
  & 3 & 3 & $0{:}1,\ 1{:}1,\ 2{:}1$ \\
\countrowsep
3 & $(1,1,1)$ & $(1,1,1)$
  & 2 & 2 & $1{:}1,\ 3{:}1$ \\
\countrowsep
3 & $(1,1,2)$ & $(1,1,1)$
  & 4 & 4 & $1{:}2,\ 2{:}1,\ 3{:}1$ \\
\countrowsep
3 & $(1,1,1)$ & $(1,1,2)$
  & 6 & 6 & $0{:}1,\ 1{:}1,\ 2{:}2,\ 3{:}1,\ 4{:}1$ \\
\countrowsep
4 & $(1,1,1,1)$ & $(1,1,1,1)$
  & 3 & 3 & $0{:}1,\ 2{:}1,\ 4{:}1$ \\
\countrowsep
4 & $(1,1,1,2)$ & $(1,1,1,1)$
  & 6 & 6 & $0{:}1,\ 1{:}1,\ 2{:}2,\ 3{:}1,\ 4{:}1$ \\
\countrowsep
4 & $(1,1,2,2)$ & $(1,1,2,2)$
  & 14 & 14 & $0{:}2,\ 1{:}1,\ 2{:}4,\ 3{:}2,\ 4{:}3,\ 5{:}1,\ 6{:}1$ \\
\countrowsep
5 & $(1,1,1,1,1)$ & $(1,1,1,1,1)$
  & 3 & 3 & $1{:}1,\ 3{:}1,\ 5{:}1$ \\
\countrowsep
5 & $(1,1,1,1,2)$ & $(1,1,1,1,1)$
  & 7 & 7 & $1{:}2,\ 2{:}1,\ 3{:}2,\ 4{:}1,\ 5{:}1$ \\
\countrowsep
5 & $(1,1,1,2,2)$ & $(1,1,1,2,2)$
  & 20 & 20 & $1{:}4,\ 2{:}3,\ 3{:}5,\ 4{:}3,\ 5{:}3,\ 6{:}1,\ 7{:}1$ \\
\countrowsep
6 & $(1,1,1,1,1,1)$ & $(1,1,1,1,1,1)$
  & 4 & 4 & $0{:}1,\ 2{:}1,\ 4{:}1,\ 6{:}1$ \\
\countrowsep
6 & $(1,1,1,1,1,1)$ & $(1,1,1,1,2,2)$
  & 29 & 29 & $0{:}3,\ 1{:}2,\ 2{:}6,\ 3{:}4,\ 4{:}6,\ 5{:}3,\ 6{:}3,\ 7{:}1,\ 8{:}1$ \\
\countrowsep
6 & $(1,1,1,1,1,1)$ & $(1,1,1,2,2,2)$
  & 40 & 40 & $0{:}1,\ 1{:}5,\ 2{:}5,\ 3{:}8,\ 4{:}6,\ 5{:}6,\ 6{:}4,\ 7{:}3,\ 8{:}1,\ 9{:}1$ \\
\countrowsep
6 & $(1,1,1,1,1,1)$ & $(1,1,1,3,3,3)$
  & 76 & 76 & \shortstack[l]{$0{:}3,\ 1{:}4,\ 2{:}8,\ 3{:}9,\ 4{:}11,\ 5{:}9,\ 6{:}10,$\\
                              $7{:}7,\ 8{:}6,\ 9{:}4,\ 10{:}3,\ 11{:}1,\ 12{:}1$} \\
\bottomrule
\end{tabular}
\endgroup
\end{table*}

\subsection{Worked example}

We use an $N=6$ example to demonstrate how the YE3T decomposition works. We define the content of the tensor product as $\boldsymbol{\nu}^\circ =[(1,1),(1,1),(1,1),(1,3),(1,3),(1,3)]$ with the repeated channel stabilizer of $S_3 \times S_3$.
Table~\ref{tab:main_rank6_parent_multiplicities} reports the sum of multiplicities $\alpha_{\boldsymbol\nu}^{\lambda L}$ for each parent $[\lambda]\otimes V_L$.
This also defines the basis size for different joint rotation/permutation representations.
The fully symmetric row $\lambda=(6)$ corresponds to the ACE sector, while $\lambda=(1^6)$ is the fully antisymmetric one.
In the Supplemental Material, these multiplicity contributions are broken down by $c_{\boldsymbol\kappa}^{\lambda}$, $d_{\boldsymbol\kappa\boldsymbol\Lambda}$, and $M_{\boldsymbol\Lambda}^{L}$.
While there are no prior comparisons for joint rotation/permutation-equivariant decompositions on this scale, we can compare the ACE-sector (permutation-invariant) decomposition space.
This is done in Table~\ref{tab:main_gepi_count_comparison}, which compares $\lambda=(N)$ multiplicities with GE-PI \cite{barthelemy_efficient_2026}.
Table~\ref{tab:main_rank6_parent_multiplicities} highlights this comparison in the $\lambda=(6) \cong \ydiagram{6}$ row; we use Young diagrams to illustrate the parent permutation state clearly \cite{fulton_representation_2013}.

\begin{table*}[t]
\color{black}
\centering
\scriptsize
\ytableausetup{boxsize=0.45em}
\setlength{\tabcolsep}{2.2pt}
\renewcommand{\arraystretch}{0.96}
\caption{
All YE3T-based $S_N\times SO(3)$ decomposition multiplicities for a tensor product with all identical non-angular indices and $\boldsymbol l=(1,1,1,3,3,3)$ for which $G_{\boldsymbol\nu}=S_3\times S_3$.
The leftmost $\lambda$ column is the parent permutation-representation carrier index (labeled by a Young diagram), and the second column gives the dimension of that permutation representation. The intervening numbered columns give $\alpha_{\boldsymbol\nu}^{\lambda L}$ for each $L$. The second-to-last column gives the total irreducible multiplicity over all $L$ in a given $\lambda$ carrier (with the full ACE-basis count in green and the purely antisymmetric-basis count in red), and the final column gives the total number of coupled components, $D_\lambda=\dim[\lambda]\sum_L(2L+1)\alpha_{\boldsymbol\nu}^{\lambda L}$.}
\label{tab:main_rank6_parent_multiplicities}
\resizebox{\linewidth}{!}{%
\begingroup\color{black}
\begin{tabular}{crrrrrrrrrrrrrrrr}
\toprule
$\lambda$ & $\dim[\lambda]$ & \multicolumn{13}{c}{$\alpha_{\boldsymbol\nu}^{\lambda L}$} & $m_\lambda$ & $D_\lambda$ \\
\cmidrule(lr){3-15}
 & & $0$ & $1$ & $2$ & $3$ & $4$ & $5$ & $6$ & $7$ & $8$ & $9$ & $10$ & $11$ & $12$ & & \\
\midrule
\rowcolor{green!8}
$\ydiagram{6}$ & 1 & 3 & 4 & 8 & 9 & 11 & 9 & 10 & 7 & 6 & 4 & 3 & 1 & 1 & 76 & 840 \\
$\ydiagram{5,1}$ & 5 & 7 & 17 & 29 & 33 & 37 & 34 & 30 & 23 & 18 & 11 & 7 & 3 & 1 & 250 & 13160 \\
$\ydiagram{4,2}$ & 9 & 10 & 25 & 41 & 48 & 52 & 47 & 41 & 31 & 23 & 14 & 8 & 3 & 1 & 344 & 31752 \\
$\ydiagram{4,1,1}$ & 10 & 8 & 27 & 38 & 49 & 48 & 45 & 36 & 28 & 18 & 12 & 5 & 2 & 0 & 316 & 31220 \\
$\ydiagram{3,3}$ & 5 & 6 & 12 & 20 & 24 & 26 & 22 & 21 & 15 & 11 & 7 & 4 & 1 & 1 & 170 & 8680 \\
$\ydiagram{3,2,1}$ & 16 & 11 & 34 & 53 & 62 & 64 & 58 & 46 & 35 & 24 & 13 & 6 & 2 & 0 & 408 & 63616 \\
$\ydiagram{3,1,1,1}$ & 10 & 5 & 19 & 25 & 33 & 30 & 27 & 20 & 15 & 8 & 5 & 1 & 0 & 0 & 188 & 17220 \\
$\ydiagram{2,2,2}$ & 5 & 4 & 8 & 13 & 16 & 16 & 13 & 12 & 8 & 5 & 3 & 1 & 0 & 0 & 99 & 4655 \\
$\ydiagram{2,2,1,1}$ & 9 & 5 & 13 & 21 & 24 & 24 & 20 & 16 & 11 & 7 & 3 & 1 & 0 & 0 & 145 & 11907 \\
$\ydiagram{2,1,1,1,1}$ & 5 & 2 & 5 & 9 & 9 & 9 & 7 & 5 & 3 & 2 & 0 & 0 & 0 & 0 & 51 & 2135 \\
\rowcolor{red!8}
$\ydiagram{1,1,1,1,1,1}$ & 1 & 1 & 0 & 1 & 1 & 1 & 0 & 1 & 0 & 0 & 0 & 0 & 0 & 0 & 5 & 35 \\
\bottomrule
\end{tabular}
\endgroup%
}
\end{table*}

\subsection{YE3T message passing models for fermion operator learning}

\begin{figure}[htbp]
\color{black}
\centering
\includegraphics[width=0.92\linewidth]{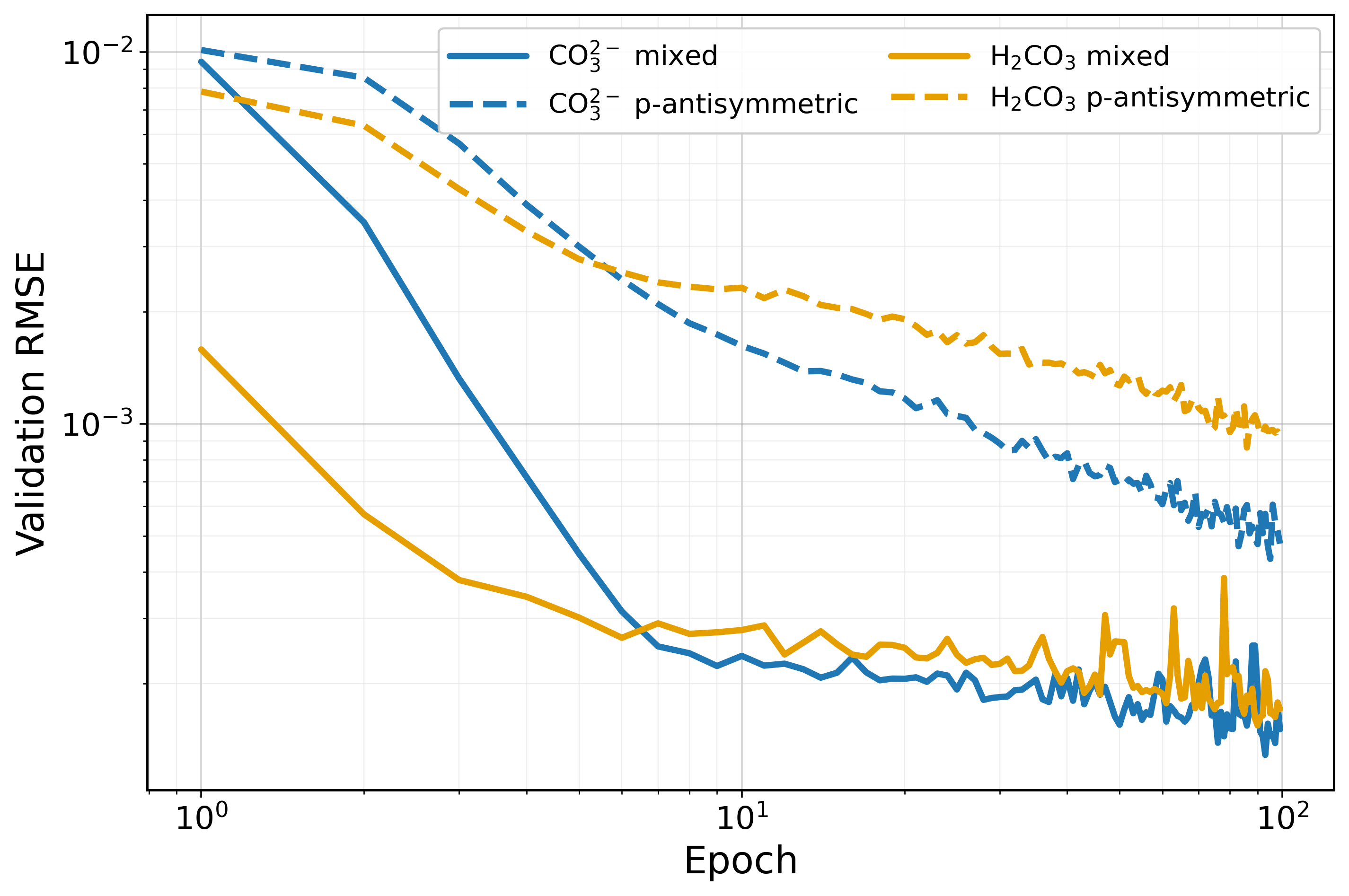}
\caption{Relative operator-error loss curves for different YE3T message-passing models applied to ERI operator learning for $H_2CO_3$ and $CO_3^{2-}$. The pair-antisymmetric models allow only $(2,2)$ character throughout the network, including the hidden layers, whereas mixed models may have varied permutation character but require $(2,2)$ symmetry at readout.}
\label{fig:eri_curves}
\end{figure}

By learning ERI matrices, we demonstrated that YE3T can provide learning advantages for electronic/fermionic problems with nontrivial permutation symmetries.
The case of $H_2CO_3$ and $CO_3^{2-}$ is a molecular example where molecular symmetries are perturbed by the addition or removal of protons.
While we do not converge the models to chemical accuracy, we are able to demonstrate that mixed permutation symmetry offers a learning advantage over models restricted to pair-antisymmetric symmetry throughout.
Figure~\ref{fig:eri_curves} reports the first 100 epochs of the mixed and pair-antisymmetric models for both $H_2CO_3$ and $CO_3^{2-}$.
Mixed permutation symmetry learned the geometry-dependent ERI matrix more quickly than restricted-permutation-symmetry models, despite both mixed and pair-antisymmetric models having the required $(2,2)$ pair antisymmetry at output.
The final validation RMSE values were $0.000129$ for the mixed-character $CO_3^{2-}$ model and $0.000529$ for the fixed pair-antisymmetric model.
For $H_2CO_3$, the corresponding values were $0.000163$ and $0.000954$.
As shown in Fig.~\ref{fig:eri_curves}, the mixed-character models plateau at lower validation RMSE values within the 100-epoch window.

\subsection{Linear YE3T model comparison}
The linear YE3T models use either a tagged basis (defined in the Supplemental Material), $u_i=\mathcal{T}_i$, or the density basis, $u_i=A_i$. For comparison with PACE, the YE3T and PACE evaluators agree to machine precision; examples are available in the LAMMPS repository \cite{lysogorskiy2021performant}.
Linear models were produced for the cost comparison datasets featured in Ref.~\citenum{zuo2020performance}.
YE3T-mixed models have mixed nontrivial intermediate rotation/permutation states, e.g., $\boldsymbol{\kappa}=[(2,2),(2,2)]$ and $\boldsymbol{\Lambda}=(2,2)$, but the final parent permutation state is symmetric, $\lambda=(8)$, and the parent rotation state is scalar, $L=0$, such that the energy models are both rotation- and permutation-invariant.

\begin{figure}[htbp]
\color{black}
\centering
\includegraphics[width=0.92\linewidth]{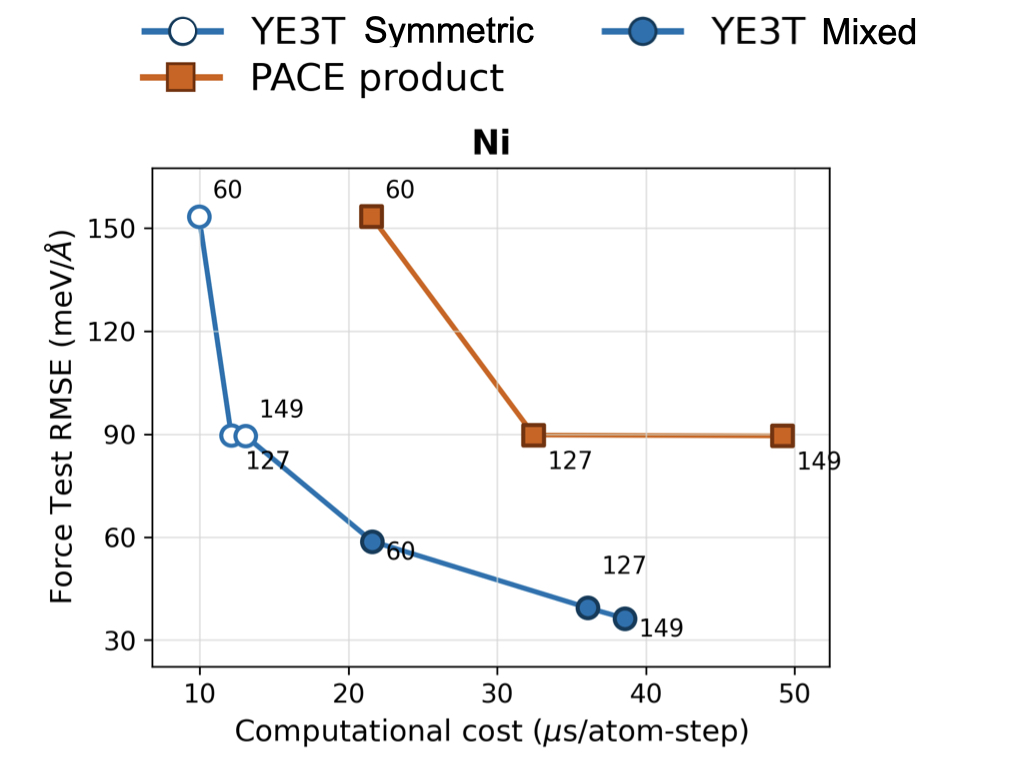}
\caption{Pareto curve for linear machine-learned potentials of Ni. YE3T-symmetric (open blue circles) and PACE product (closed orange squares) are separate but equivalent implementations of the same ACE model. They are permutation-invariant throughout.
The YE3T mixed linear models have permutation-equivariant intermediates and are rotation/permutation invariant outputs (closed blue circles).
Numbers of descriptors per model are annotated next to plot markers.
YE3T methods are faster, more accurate, and scale better with the number of features.
The x-axis shows LAMMPS CPU loop time in microseconds per atom-step. The y-axis is test-set force RMSE, using the same test set as Ref.~\citenum{zuo2020performance}.}
\label{fig:linear_pareto}
\end{figure}

Linear models were trained with $L_2$-norm ridge regression and a small ridge penalty of $1.0\times10^{-8}$ for Li, Mo, Cu, Ni, Si, and Ge.
For each system, hyperparameters such as the radial cutoff and radial-decay parameters were optimized using DAKOTA 6.16's efficient global optimizer \cite{adams_dakota_2022}.
Models were optimized such that they run stable molecular dynamics in LAMMPS for 10,000 time steps, but other properties were not optimized.
Final optimal hyperparameter values were recorded and used in the LAMMPS potential files.
The potential files record the choices of $n_{\max}$ and $l_{\max}$ for each descriptor rank; the models use ranks 1--8 with $n_{\max}=6$ and $l_{\max}=4$.
The reported LAMMPS timings use one CPU thread on an Intel Core i9 processor.
The general linear YE3T models with mixed rotation/permutation intermediates and purely symmetric ACE formalisms do not allow for an exact match of $\eta,l$ descriptor lists, but they are constructed such that descriptor counts are similar.
The descriptor count per model is plotted in Fig.~\ref{fig:linear_pareto}.

The results in Fig.~\ref{fig:linear_pareto} are compelling, as YE3T provides a Pareto advantage over existing linear PACE models (orange squares).
This includes a large speedup provided by YE3T-symmetric (open blue circles) models that exactly reproduce ACE models.
Linear YE3T models with mixed nontrivial permutation intermediates (labeled as ``YE3T-mix'' in the figure as closed blue circles) offer higher accuracy with fewer descriptors. The YE3T reformulations of $\lambda=(N)$ ACE models use Schur--Weyl-adapted kernels described in the Supplemental Material, reproduce ACE evaluators to numerical precision, and are 2--5 times faster even for small models. These are labeled ``YE3T-symmetric.''
Between these models and the old PACE product evaluator, labeled ``PACE product'' in the plot, the $2$--$5\times$ speedup is demonstrated for various descriptor counts.
One compelling advance provided by YE3T is that adding more features can come at little extra cost; YE3T has much better scaling with the number of descriptors. Nearly negligible increases in computational cost per time step for 127- versus 149-descriptor models were found with YE3T, whereas ACE nearly doubles the computational cost for the same added features.
There is also an approximately $2\times$ speed advantage over the PACE recursive evaluator in LAMMPS, as we demonstrate in the Supplemental Material with strong-scaling results. As reported in the Supplemental Material, the YE3T speed advantage grows for larger ranks (e.g., $N=16,32,64$).

The YE3T-symmetric Pareto curve for MLIPs without permutation-adapted intermediates also demonstrates a key limitation of the ACE basis and a clear advantage of YE3T.
The ACE model errors saturate at certain values and do not continue to improve as more ACE features are added.
The YE3T models saturate later than ACE models do in this Ni model, as well as in five other models reported in the Appendix.
In fact, with these small models, it is not clear that the YE3T model errors have been saturated.
Adapting the tensor/cluster basis to \textit{all} the physical symmetries in the system (rotations and permutations), rather than only rotations, offers measurable benefits. However, further exploration of model-plateau limitations is left for future work.

Overall, nontrivial permutation information can be included systematically to improve accuracy, as demonstrated with adjusted descriptor counts for each element, but at increased computational cost.
These advantages remain even when using the ``recursive evaluator'' in LAMMPS for ACE features, as YE3T still significantly outperforms that evaluator as well.
Evidence for this is given for models with various tensor product ranks in the Supplemental Material. This advantage is the practical result of the Schur--Weyl decomposition versus the brute-force tensor product illustrated in Fig.~\ref{fig:main_cg_vs_ye3t}.
One compelling result of YE3T evaluators and kernels is that additional YE3T features may be added at minimal additional cost while still improving models.
This is evident for the YE3T-symmetric linear models (open blue circles in Fig.~\ref{fig:linear_pareto}).
These permutation-invariant YE3T models reproduce PACE models exactly. They are YE3T reformulations evaluated with symmetry-adapted kernels.
Additional descriptors may be added at little cost by using mixed/nontrivial permutation-character intermediates (the YE3T-mixed closed blue circles in Fig.~\ref{fig:linear_pareto}).
The YE3T models also provide higher accuracy with fewer descriptors.
In the Supplemental Material, LAMMPS strong-scaling results demonstrate that YE3T advantages remain over PACE recursive evaluators and across element types.

Similar Pareto advantages with ACE over previous methods were reported in Drautz 2019 \cite{drautz_atomic_2019}.
At a high level, these advantages can be attributed to the efficient tensor product basis for $SO(3)$ product carriers \cite{dusson_atomic_2022}. This is implicitly supported by the development of MACE and GRACE, which rely on the rotation-equivariant machinery of ACE \cite{batatia_mace_2022,bochkarev_graph_2024}.
Analogous advantages were observed with YE3T rotation/permutation equivariance as it was extended to rotation/permutation-equivariant message-passing models, including the ERI message-passing application.

\section{Conclusions}

Young-E(3) tensor (YE3T) equivariant methods were developed that simultaneously treat rotation and permutation symmetries.
YE3T was connected to traditional cluster-expansion methods as well as the atomic cluster expansion.
Rather than symmetrizing with respect to permutations prior to symmetry adaptation to rotations, as is done in ACE, YE3T treats both simultaneously and does not lose physical permutation information due to premature averaging or suffer from linearly dependent descriptors reported in early ACE works.
We are able to derive a complete, independent ACE basis through YE3T methods that avoids an overcomplete construction entirely and is complete and orthonormal by construction (for a given fixed $N$ and tensor product content).
These bases are derived from a representation-theoretical standpoint that enables faster computation and more rigorously defined bases. The joint rotation/permutation carrier spaces were defined and decomposed using repeated application of Schur--Weyl decomposition.
This produces a multiplicity space with exact basis coordinates and allows for derivation of rotation/permutation coupling coefficients.
In practice, this helps produce faster kernels for existing $S_N\times SO(3)$ problems like ACE, where we show a significant speedup over both the recursive and product evaluators in LAMMPS.
The YE3T theory also extends beyond permutation-invariant models to implement fast rotation- and permutation-equivariant models.

We validate YE3T against existing methods such as the GE-PI method from recent work.
This shows that we are able to produce a complete, independent, permutation-invariant, rotationally equivariant basis (e.g., that used for ACE/MACE).
The symmetry-adapted kernels in YE3T provide widespread construction-time advantages across the reported label set, including speedups of up to $1000\times$ relative to cuEquivariance \cite{rylieweaver9_accelerate_2024}.

Through YE3T, we push beyond E(3) equivariance and linear ACE models that carry intermediate rotation information but no permutation-equivariant information.
We formulate the architecture for rotation/permutation-equivariant message-passing networks with hidden states that carry permutation information in addition to rotation information.
This is the message-passing analog of linear YE3T, just as MACE, GRACE, and related methods are message-passing analogs of linear ACE.

Furthermore, YE3T enables more general joint rotation/permutation-equivariant modeling beyond atomistic and electronic systems. Future work could apply this to other particle systems and extend it to other applications that rely on E(3) equivariance and also carry permutation information. With machinery and theory analogous to e3nn, YE3T theory provides a means to perform general rotation/permutation-equivariant message passing in a controlled, systematic way.

\section{Acknowledgments}
James Goff thanks Ralf Drautz for helpful conversations about combining non-angular indices in ACE tensor product representations.

James Goff thanks Jigyasa Nigam for helpful conversations about nonlinear dependencies between ACE-like tensor products.

The efforts in this work were supported by the Laboratory Directed Research and Development program at Sandia National Laboratories, a multimission laboratory managed and operated by National Technology and Engineering Solutions of Sandia LLC, a wholly owned subsidiary of Honeywell International Inc. for the U.S. Department of Energy’s National Nuclear Security Administration under contract DE-NA0003525. This paper describes objective technical results and analysis. Any subjective views or opinions that might be expressed in the paper do not necessarily represent the views of the U.S. Department of Energy or the United States Government.

\section{AI disclosure}
ChatGPT 5.6 was used for curating citations, generating TikZ illustrations, and table structuring/formatting.
Claude Code Fable on max reasoning was used for LAMMPS integration and kernel optimization of YE3T through \texttt{pair\_style ye3t}.

\section{Data Availability}
The LAMMPS examples are available at \url{https://github.com/ye3t-equivariance/ye3t-lammps}.
The YE3T software is available at \url{https://github.com/ye3t-equivariance/ye3t}.
Other data will be available upon reasonable request.
\appendix
\section{Additional performance results}

The Pareto advantages offered by YE3T models over ACE are observed across multiple elements.
\begin{figure}[htbp]
\color{black}
\centering
\includegraphics[width=0.92\linewidth]{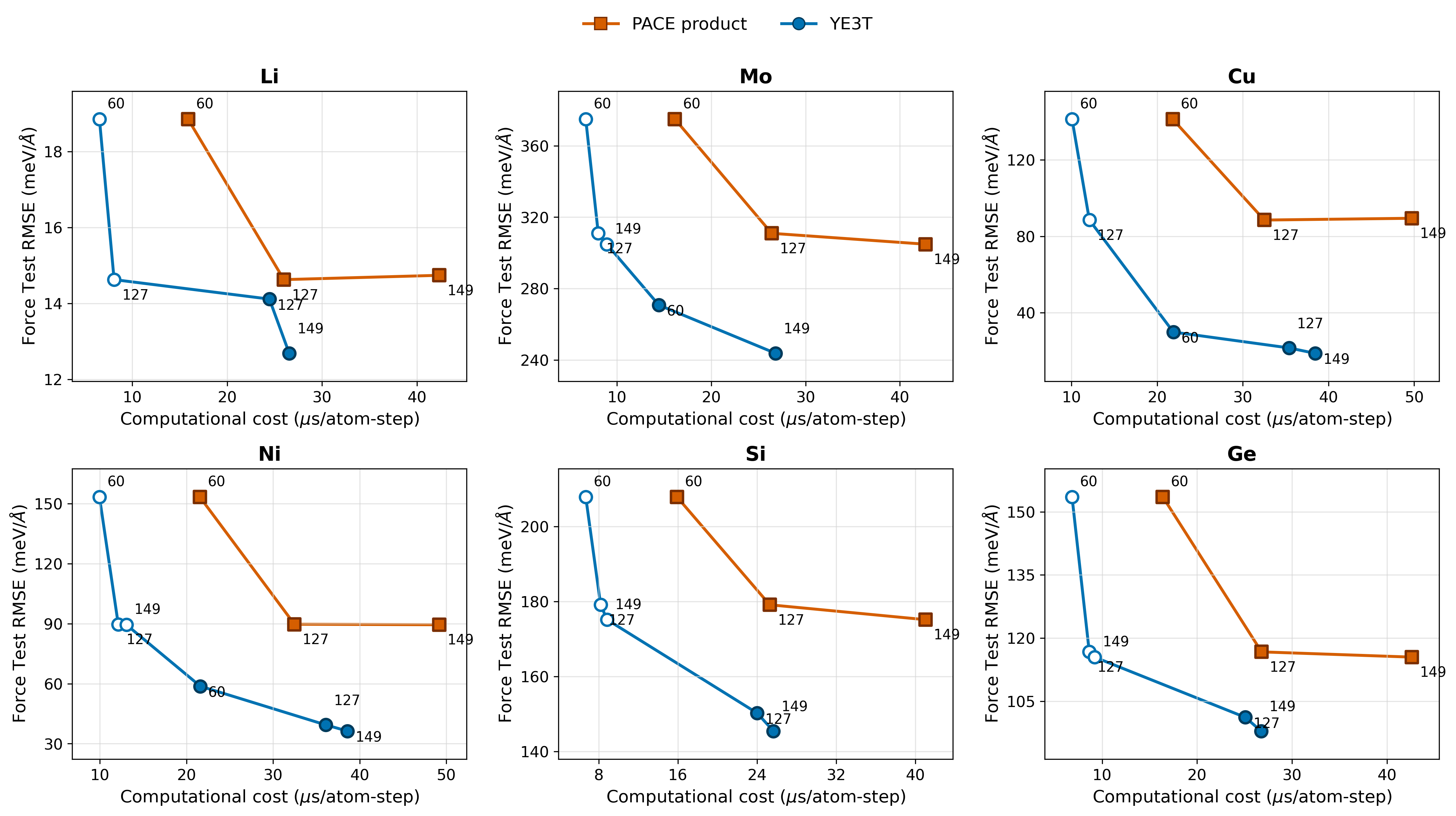}
\caption{For linear models, the force RMSE Pareto plots are provided. The x-axis shows LAMMPS CPU loop time in microseconds per atom-step (smaller is better). This uses the same validation set as Ref.~\citenum{zuo2020performance}.}
\label{fig:all_linear_pareto}
\end{figure}

The time to build a YE3T-symmetric basis (and coupling coefficients) is the same task pursued by many ACE and MACE-facing methods and libraries. Specifically, the coupling of $V_l\otimes V_l\otimes V_l\otimes\cdots$ to $V_L$ through the generalized Clebsch--Gordan coupling.
With the representation-theoretical foundations developed here, we can couple from $V_l\otimes V_l\otimes V_l\otimes\cdots$ to $V_L$ while ensuring that the coupling paths correspond to orthogonal basis functions.
Given the symmetry-adapted kernels developed here, we obtain widespread speedups across the reported label set, including cases up to $1000\times$ faster than cuEquivariance \cite{rylieweaver9_accelerate_2024}.

\begin{figure*}[t]
\color{black}
\centering
\includegraphics[width=\textwidth]{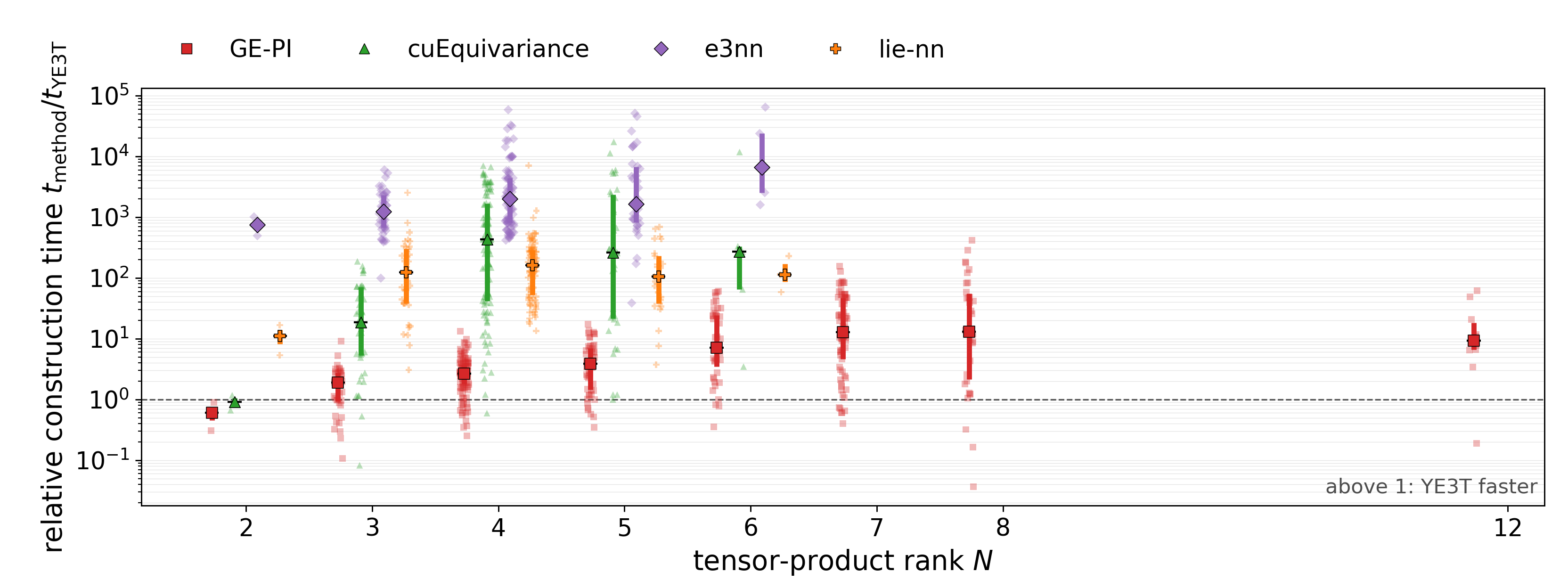}
\caption{Construction-time comparison for basis couplings in the $\lambda=(N)$ permutation-invariant sector used in ACE. This is reported for a list of 374 $\boldsymbol{\eta},\boldsymbol{l}$ combinations at ranks $N=2$--$8$ and $12$, detailed in the SI.
The figure reports $t_{\mathrm{method}}/t_{\mathrm{YE3T}}$, or the speedup YE3T provides over other methods, for completed pairs with matching output multiplicities. Translucent markers show individual labels; opaque markers and vertical bars show the median and interquartile range. Completion and failure accounting for the full basis-label list is given in the SI; unsuccessful or unrecorded cases are not assigned timing values. Each attempt used a 600~s timeout, and cuEquivariance dense arrays were skipped above an estimated 2~GiB \cite{rylieweaver9_accelerate_2024}.}
\label{fig:main_external_basis_construction}
\end{figure*}

\bibliography{references}

\end{document}